\documentclass{mn2e}
\input epsf.sty

\def\Rg{r_{\rm g}}
\def\lh{l_{\rm h}}
\def\lha{l_{\rm h,1}}
\def\lhb{l_{\rm h,2}}

\def\lhmax{l_{\rm h, max}}
\def\lsoft{l_{\rm s}}
\def\ls0{l_{\rm s,0}}
\def\tdyn{t_{\rm dyn}}
\def\fbreak{f_{\rm break}}

\def\tmin{\tau_{\rm min}}
\def\tmax{\tau_{\rm max}}
\def\taur{\tau_{\rm r}}
\def\taud{\tau_{\rm d}}
\def\rmin{r_{\rm min}}
\def\rmax{r_{\rm max}}
\def\OmegaK{\Omega_{\rm K}}
\def\taues{\tau_{\rm es}}
\def\tauhot{\tau_{\rm hot}}
\def\Rflare{R_{\rm flare}}
\def\Lflare{L_{\rm flare}}
\def\Fdisc{F_{\rm disc}}
\def\Fillum{F_{\rm irr}}
\def\Ftot{F_{\rm tot}}
\def\Ltot{L_{\rm tot}}
\def\sigmaT{\sigma_{\rm T}}
\def\me{m_{\rm e}}

\def\kT{k T_{\rm e}}
\def\Mdot{{\dot M}}
\def\MSun{{\rm M}_{\odot}}
\def\MEdd{{\dot M}_{\rm Edd}}
\def\LEdd{L_{\rm Edd}}

\def\kapabs{\kappa_{\rm abs}}
\def\kapes{\kappa_{\rm es}}
\def\thetae{\theta_{\rm e}}
\def\ptrans{p_{\rm tr}}
\def\prefl{p_{\rm rf}}
\def\Sref{S_{\rm refl}}
\def\Sprim{S_{\rm prim}}
\def\Ka{K$_\alpha$\ }

\title[Fourier-resolved spectroscopy]
{Predictions for Fourier-resolved X-ray spectroscopy from the model of
magnetic flare avalanches above an accretion disc with hot ionized skin}

\author[P. T. \.{Z}ycki]{Piotr T. \.{Z}ycki\thanks{e-mail: ptz@camk.edu.pl} \\
    Nicolaus Copernicus Astronomical Center, Bartycka 18, 00-716 Warsaw,
Poland}

\date{20 Feb 2002}

\begin{document}
\label{firstpage}

\maketitle

\begin{abstract}

The magnetic flare avalanches model of Poutanen \& Fabian for
X--ray variability of accreting black holes 
is combined with computations of vertical structure of illuminated accretion
discs in hydrostatic equilibrium. The latter predict the existence of a
hot ionized skin, due to the thermal instability of X--ray illuminated
plasma. The presence of such ionized skin, with properties dependent on
disc radius, introduces a dependence of the emitted X--ray spectrum
on the position on the disc. If the position is related to the time scale
of the flares, the X--ray energy spectra (both the primary continuum
and the reprocessed component) gain an additional dependence on Fourier 
frequency, beside that resulting from spectral evolution during a flare.
We compute the Fourier frequency resolved spectra in this model and 
demonstrate that the presence of the hot skin introduces trends opposite
to those observed in black hole binaries. Furthermore, the flare profile is 
strongly constrained, if the Fourier frequency dependence due to spectral 
evolution is to agree with observations.

\end{abstract}


\section{Introduction}
 \label{sec:intro}

X--ray emission from accreting compact objects carries information about
geometry and physical conditions in the immediate vicinity of central
objects. Unfortunately, the information is ``encrypted'' into a series
of numbers, each number (event) corresponding to a detection of a photon from 
the source. Two main characteristics of each event are the photon energy
and arrival time (we neglect here polarization of radiation).
One way of analyzing the data stream is then to project the events onto 
the energy axis, thus
suppressing the timing information. Thus obtained energy spectra give
a first, rough insight into physical mechanism of emission of radiation.
For example, observations of cutoffs at $\sim 100$ keV in low/hard
state of accreting black holes demonstrate that
the Comptonizing electrons have a thermal rather than non-thermal distribution.
From modelling the spectra a mean plasma temperature and scattering optical 
depth can be inferred. However, any possible distributions of plasma temperature
and density cannot be recovered from time-averaged spectra
(e.g.\ Kazanas, Hua \& Titarchuk 1997; Coppi 1999).

In an analogous way, projecting the data onto the time axis produces
a light curve, which can be analyzed either in the time or Fourier domain.
The results, too, enable determination of certain average properties.
For example from power spectral density (PSD), the root mean square 
variability or constraints on distribution of shots' time scales in the shot 
noise model can be found (Lehto 1989; Lochner, Swank \& Szymkowiak 1991).


Further progress can be made by including at least limited information from
along the other parameter axis. For example, cross-correlation analysis
(both in time and Fourier domain)  reveal that higher energy X--rays
lag the lower energy ones (Miyamoto \& Kitamoto 1989), which one can
attempt to interpret as e.g.\ spectral evolution during a shot (flare).
Then the energy dependent auto-correlation functions give average shot 
profiles as a function of energy (see review in Poutanen 2001 and 
references therein). 

Similarly, studying a sequence of energy spectra, each one averaged over 
a certain short time interval, enables determination of not only the mean 
spectral properties, but also ``higher orders'' statistics properties. 
The most important result that has emerged recently is that 
spectral properties of a given source vary with time, but 
correlations between parameters exist. The power law slope, $\Gamma$, 
is correlated with the amplitude of reflection $R$ and the width
of the Fe \Ka line (Zdziarski, Lubi\'{n}ski \& Smith 1999; 
Gilfanov, Churazov \& Revnivtsev 1999; Revnivtsev, Gilfanov \& Churazov 2001; 
\.{Z}ycki, Done \& Smith 1998). Interestingly, the same correlations
seem to exist between sources in a sample (Zdziarski et al.\ 1999; 
Lubi\'{n}ski \& Zdziarski 2001).

A number of geometrical/physical scenarios can be built based on  
the $R$--$\Gamma$ correlation alone.
The crucial element of all models is the feedback between the hot, X--ray
emitting plasma and the cooler, optically thick plasma: most of the soft 
photons crossing the hot plasma have to come from thermalization/reprocessing 
of the illuminating hard X--rays (Poutanen 1998 and references therein). 
Given this condition alone,
models can be constructed belonging to two main classes,
(1) with the standard accretion disc disrupted at a certain radius,
and (2) with the standard accretion disc extending to the last stable
orbit:
\begin{itemize}
 \item  standard accretion disc disrupted at a given radius, and filled with
       hot, optically thin, X--ray producing flow (Esin, McClintock \& Narayan
      1997; Poutanen, Krolik \& Ryde 1997;
        R\'{o}\.{z}a\'{n}ska \& Czerny 2000), with the amount
      of overlap between the two phases controlling the correlations,
 \item hot, optically thin, quasi-spherical flow with cold cloudlets
    inside (Rees 1987; Kuncic, Celotti \& Rees 1997), with the cloudlets 
     filling factor controlling the correlations,
 \item spherical accretion of cold clouds  onto a central X--ray
  source (Collin-Souffrin et al.\ 1996; Malzac 2001), 
 with the covering factor of the clouds as the  control parameter,
 \item mildly relativistic outflow of the X--ray emitting plasma, above
    a standard accretion disc, with the outflow velocity controlling
    the correlations (Beloborodov 1999a,b; Janiuk, Czerny \& \.{Z}ycki 2000;
   Malzac, Beloborodov, Poutanen 2001),
 \item hot, ionized skin on top of the accreting disc, due to thermal
    instability of the X--ray illuminated plasma, with the skin thickness
    as the control parameter (R\'{o}\.{z}a\'{n}ska \& Czerny 1996;
    Nayakshin, Kazanas \& Kallman 2000; \.{Z}ycki \& R\'{o}\.{z}a\'{n}ska 
    2001).
\end{itemize}

Also, models for time variability predicting correct PSD, time lags 
and coherence function can be constructed for the two main 
geometries (Poutanen \& Fabian 1999, hereafter PF99; 
B\"{o}ttcher \& Liang 1999; review in Poutanen 2001).
 Therefore, 
based solely on the limited spectral/temporal data distinguishing between
these scenarios appears very difficult, if at all possible
(see e.g.\ discussion and application to Cyg X-1 in Di Salvo et al.\ 2001).
Employing more sophisticated tools, involving both spectral
and temporal properties is clearly required.

The Fourier frequency resolved ($f$-resolved hereafter) energy spectra seem to 
be a potentially powerful tool to apply to the problem of distinguishing 
the models. 
The $f$-resolved spectra are energy spectra in the limited range of Fourier 
frequency and so, roughly speaking, represent spectra at different time scales
of variability (Revnivtsev, Gilfanov \& Churazov 1999; Gilfanov, Churazov 
\& Revnivtsev 2000a,b). The $f$-resolved spectra found from data of X-ray 
binaries show a number of features and correlations. Most
importantly, the $R$-$\Gamma$ correlation is fulfilled by the $f$-resolved
spectra in the sense that the higher the Fourier frequency, the
harder the spectrum, and the smaller the amplitude of reflection
(Revnivtsev et al.\ 1999).
This result indicates that the same mechanism which is responsible for the
correlation on long timescales ($\sim 10^3$ sec $\approx 10^5$--$10^6\tdyn$,
$\tdyn=\OmegaK^{-1}$ being the dynamical timescale), acts also on much
shorter time scales, at least down do $\sim 0.01$ sec, the time bin of
the Revnivtsev et al.\ analysis. This fact provides the motivation for
our attempt to connect the spectral and timing models and predict 
their combined results.
Therefore in this paper we consider  a simple combination of the 
model  of PF99, with the structure of illuminated 
accretion discs in hydrostatic equilibrium. The former is the most
complete X--ray emission model so far, since it addresses both
the spectral and timing properties of Comptonization emission. 
The latter naturally predicts
the appearance of a hot, ionized skin on top of the disc (R\'{o}\.{z}a\'{n}ska
\& Czerny 1996; Nayakshin et al.\ 2000), as a result
of the thermal instability of the X-ray irradiated plasma (Field 1965;
Krolik, McKee \& Tarter 1981). The ionized skin
reduces the effectiveness of reprocessing and thermalization of the hard 
X--rays. The flux of soft photons is reduced, which may lead to a harder 
continuum spectrum.
If the ionized skin is thickest close to the black hole (Nayakshin 2000;
\.{Z}ycki \& R\'{o}\.{z}a\'{n}ska 2001), then the hardest spectra, with lowest
reflection amplitude are expected to originate there. If the short time scale
(high Fourier frequency, $f$) flares originate close to the black hole, then
one may expect the high-$f$ spectra to be harder, and show less reflection
compared to low-$f$ spectra, precisely as observed from accreting black hole 
binaries.

The plan of the paper is as follows: 
in Section~\ref{sec:model} we define the model, in Section~\ref{sec:test}
we  perform some test
computations to gain deeper insight into the properties of frequency-resolved
spectra, Section~\ref{sec:runs} contains results of full model calculations
and their comparison with data; the results are discussed in 
Section~\ref{sec:discuss}.

\section{The Model}
\label{sec:model}

 \subsection{Time variability}
 \label{sec:tvaria}

We envision that the X--rays are produced in flares occurring above an 
accretion disc. The variability part of our model is therefore
identical to that of PF99 (see also Stern \& Svensson 1996), so only a brief 
description is given here for completeness. There is a number of spontaneous 
flares, $\lambda$ per second, each of which can give rise to an avalanche
of stimulated flares. The number of stimulated flares is drawn from a 
Poissonian distribution with the mean $\mu$. Each stimulated flare is 
delayed from the stimulating flare by an interval $\Delta t$, drawn
from a Poissonian distribution of mean $\alpha \tau_1$, where $\tau_1$ is
the time scale of the stimulating flare while $\alpha$ is a parameter.
The time scales of all flares are drawn according to a power law
probability distribution $\propto \tau^{-p}$ between $\tmin$ and $\tmax$.

As in PF99 all flares have the same profile, which describes the heating
of electrons as a function of time, $\lh(t)$. Here,
$\lh$ is the compactness parameter, $l \equiv L\sigmaT/(\Rflare \me c^3)$,
where $L$ is the luminosity and $\Rflare$ is the radius of a flare.
The flare profile turns out to be decisive for the obtained dependence
of energy spectra on Fourier frequency, therefore we will employ two
of prescriptions:
\begin{itemize}
   \item the function used by PF99, 
\begin{equation}
\label{equ:pfheat}
 \lha(t) \propto (t/\tau)^2 \exp(-t/\tau),
\end{equation} 
   \item a double exponential function defined as
\begin{equation}
\label{equ:dbleexp}
\lhb(t) \propto \left\{
 \begin{array}{lc}
  \exp(t/\taur),  & \mbox{for $t<0$} \\
  \exp(-t/\taud), & \mbox{for $t>0$},
 \end{array}
        \right.  
\end{equation}
where $\taur$ and $\taud$ represent the rise and decline time scales
and $\taur\gg\taud$  (Maccarone, Coppi \& Poutanen 2000).
\end{itemize}
The functions are normalized
to give an assumed value of peak compactness, $\lhmax$, for which we
adopt $\lhmax=10$.  The flare decline time scale is assumed $\taud=0.1\taur$.

The flux of  soft photons through the flare is given as 
$\lsoft(t) = \ls0 + D(t)\lh(t)$ (PF99), where the second term describes the
feedback due to reprocessing of the hard X--rays. Here, our first extension
of the PF99 model is introduced: the feedback expression contains
a factor describing the effectiveness of thermalization due to the ionized
skin,
\begin{equation}
 \label{equ:feedb}
  D(t) = S(t)\, {D_0 \over 1+3 \left[H(t)\right]^2}.
\end{equation}
Here $S(t)= S[\tauhot(t)]$ is the fraction of the illuminating hard X--rays 
that get through the ionized skin to the cold disc, as a function of $\tauhot$
(equation~\ref{equ:st}), $D_0$ is a  parameter, 
while $H(t)$ is the height of the flare above the disc.

\subsection{Radial dependencies}
 \label{sec:radial}

The crucial extension of the PF99 model which needs to be made here is to 
relate 
the time scale of a flare to its radial location on the accretion disc.
We will consider two cases:
\begin{enumerate}
 \item There is a unique correspondence between the time scale of an 
  individual flare (both spontaneous and stimulated) and its location on 
  the disc. Since we do not assume any correlation between time scales of
  a stimulating and stimulated flare, this means that spatial locations
  of flares within an avalanche are not correlated.
 \item The unique correspondence between $\tau$ and $r$ holds for
  spontaneous flares, while all flares stimulated by a given spontaneous flare
  occur at the location of the latter (Merloni \& Fabian 2001).
\end{enumerate}
The essential assumption here is that there is at least partial
correspondence  between flare's time scale and its radial position.
To quantify the discussion we assume that the flare time scale has the same
 radial dependence as the inverse of the Keplerian frequency, 
$\tau\propto r^{3/2}$. Moreover, the shortest
flare time scale is assumed to correspond to the marginally stable orbit
at $6\Rg = 6 G M/c^2$. This gives
\begin{equation}
 \label{equ:rvstau}
 r = 6\Rg  \left({\tau \over \tmin}\right)^{2/3}.
\end{equation}
One consequence of such assumption is the existence of the maximum
radius of the flares-covered disc, 
$\rmax = 6\Rg \left(\tmax/\tmin\right)^{2/3} \approx 170\Rg$.

Some other parameters of the model need to be considered as functions of 
radius, too. The compactness 
\begin{equation}
\ls0 \equiv  \pi \Rflare \Fdisc {\sigmaT \over \me c^3}
\end{equation}
(see equation 1 in PF99) is a function of radius, since both the 
internal disc emission, $\Fdisc$,
and the flare's radius, $\Rflare$ may in general be functions of radius.
For the internal disc emission we adopt the standard, Newtonian formula
(Shakura \& Sunyaev 1973)
\begin{equation}
 \Fdisc(r) = \xi \Ftot (r) = \xi\,{3 \over 8 \pi} {G M \Mdot \over r^3} 
   \left(1-\sqrt{6\,\Rg\over r}\right),
\end{equation}
where $\xi$ is the fraction of total energy dissipated inside the
disc ($1-\xi$ dissipated in the active corona). 
We will assume that the radius of a flare, $\Rflare$, is 
constant with the radial position.
The value of $\Rflare$ 
will be derived from global energy budget for the hot plasma:
the total, time-integrated luminosity of all flares equals energy dissipation
in the hot phase,
\begin{equation}
   {\me c^3 \over \sigmaT} \sum_{i=1}^{N} 
  \int_{0}^{T} \,l_{{\rm h},i}(t)R_{{\rm flare},i}\, dt
      =  (1-\xi) \Ltot T,
\end{equation}
where the sum is over all flares, 
$\Ltot=\int_{6 \Rg}^{\infty} 2\pi r\Ftot(r)\,dr $ is the total accretion 
luminosity, while $T$ is the total time interval.
Note that the radial dependence of $\ls0$ introduces slight overall radial
dependence of spectra, since the time evolution of electron temperature
depends on the value of $\ls0$.

\subsection{The energy spectrum}

 \subsubsection{Comptonized continuum}

We assume that the continuum spectrum is due to Comptonization of soft
photons from the disc and most of these photons come from reprocessing
of the hard X--ray flux.
The Comptonized spectrum is computed using the {\sc thComp} code
of Zdziarski, Johnson \& Magdziarz (1996), based on solution of
the Kompaneets equation. This is parameterized by the photon index
of the spectrum, $\Gamma$ and plasma temperature, $\kT$.  These parameters
are computed using formulae from Beloborodov (1999b):
$\Gamma = 2.33 (\lh/\lsoft)^{-1/6}$, $\Gamma=(9/4) y^{-2/9}$, and 
$y=4(\thetae+4\thetae^2)\taues (\taues+1)$,
where $y$ is the Comptonization parameter and $\thetae=\kT / \me c^2$.
We assume $\taues = 1.8$ (PF99).

 \subsubsection{The reprocessed component}

\begin{figure}
 \epsfysize = 7 cm
 \epsfbox[18 220 600 700]{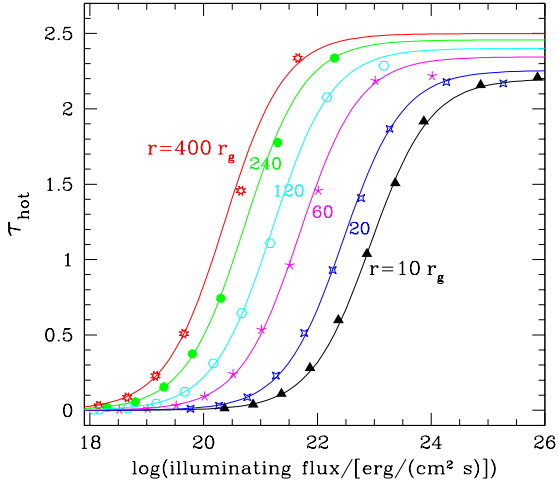}
 \caption{Thomson thickness of the hot skin as a function of the illuminating
 flux and disk radius computed using the code of \.{Z}ycki \& 
 R\'{o}\.{z}a\'{n}ska (2001). 
 Fitting formula (Eq.~\ref{equ:tauhot}) has been used as a representation
 of the results (curves) for a given value of the spectral index of
 illuminating radiation. Power law with $\Gamma=1.65$ was used in the
 presented case.
\label{fig:tauhot}}
\end{figure}

From equation~(1) of PF99 one can derive the flux of radiation illuminating
the disc in the immediate vicinity of the flare,
\begin{equation}
 \Fillum(t) = {\me c^3\over \sigmaT \pi} {\lh(t)\over \Rflare}
        {\eta D_0 \over 1+3 \left[H(t)\right]^2},
\end{equation}
where the height of the flare above of the disc is $H(t) = H_0 t/\tau$,
with $\eta=0.5$ and $D_0$ and $H_0$ as parameters. 
From the known $\Fillum$ the thickness of the ionized skin can be computed at
the considered radius ($0.25\Fillum$ is used in the actual computations,
representing spatially average flux). 
This is in general a complicated task, requiring
solving simultaneously equations of the vertical structure of the disc
and the ionized skin (R\'{o}\.{z}a\'{n}ska \& Czerny 1996; 
R\'{o}\.{z}a\'{n}ska 1999; Nayakshin et al.\ 2000; 
\.{Z}ycki \& R\'{o}\.{z}a\'{n}ska 2001). The computations cannot be 
performed in real time because of the large number
of spectra needed. Simulating a 256 sec lightcurve with $1/128$ sec
time resolution requires computing more than $10^5$ spectra. Therefore,
we use the code of \.{Z}ycki \& R\'{o}\.{z}a\'{n}ska (2001) to
generate a grid of $\tauhot$ as a function of $\Fillum$ and $r$.
A fitting formula is then used to describe the results,
\begin{equation}
 \label{equ:tauhot}
 \tauhot(\Fillum,r) = N \left\{1 +
 \tanh { \log\left[ 
 {\Fillum\over F_0} \left({r\over 20\Rg}\right)^{1.6}
 \right] \over W}\right\},
\end{equation}
where $F_0 =  10^{22.45}\ {\rm erg\ cm^{-2}\ s^{-1}}$, $W=1.05$,
 and $N=1.034+0.0937\log(r/2\Rg)$.
The accuracy of the formula is demonstrated in Fig.~\ref{fig:tauhot}.

Knowing $\tauhot$ we estimate the effective amplitude of the cold reprocessed
component as
\begin{equation}
 R = { [\ptrans(\tauhot)]^2 \over 1 + \prefl(\tauhot) },
\end{equation}
where $\ptrans(\tauhot)$ is the probability of an illuminating photon being
transmitted through the hot skin to the cold disc, while 
$\prefl(\tauhot) = 1-\ptrans(\tauhot)$ is the probability of the photon
being back-scattered in the hot skin and contributing to the primary
continuum (\.{Z}ycki \& R\'{o}\.{z}a\'{n}ska 2001). The square of $\ptrans$
in the numerator comes from the fact that  the reflected photons have to
go twice through the hot skin. The probabilities are estimated using 
a Monte Carlo code for a purely scattering atmosphere. The above
prescription for $R$ is obviously rather approximate, so we do not attempt
to e.g.\ reproduce the correct observed amplitude of the reprocessed
component. 
The function $S(\tauhot)$ describing the dependence of the soft flux
from thermalization on $\tauhot$ (see Eq.~\ref{equ:feedb})
is taken from Poutanen (2002),
\begin{equation}
 \label{equ:st}
  S(\tauhot) = (1-a){7 + \exp(-10\tauhot) \over 2 [4+3\tauhot(1-a)] },
\end{equation}
where $a$ is the cold reflection albedo, $a\approx 0.2$.

\begin{figure*}
 \epsfysize 7cm
 \epsfbox[18 480 600 700]{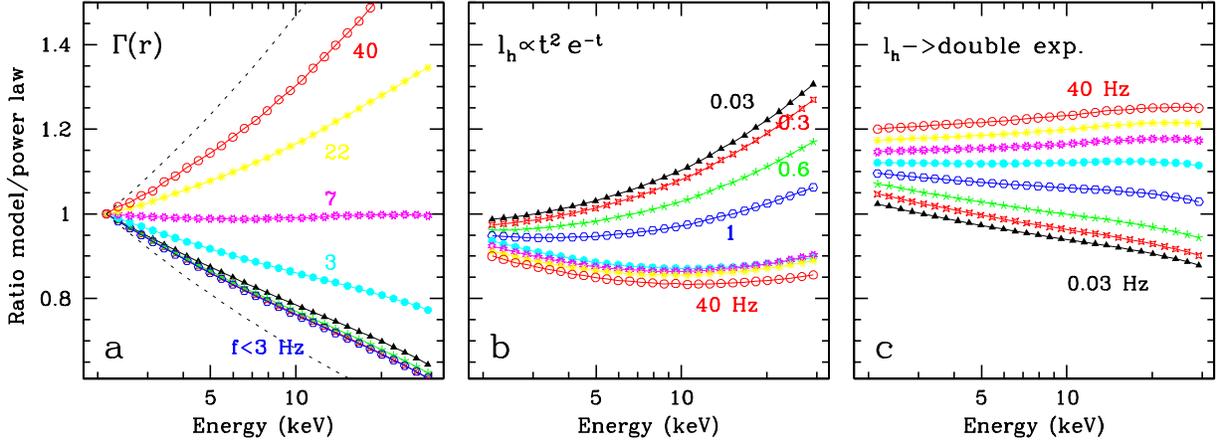}
 \caption{Test cases of Fourier-$f$ resolved spectra, plotted as ratio
  to $E^{-1.75}$ power law.
 Panel {\it (a)\/} shows an artificial case of a power law dependence of
 $\Gamma(r)$ such that $\Gamma(\rmin) = 1.5$ while $\Gamma(\rmax)=2$,
 i.e.\ the shorter flares have harder spectra.  The dotted lines show the 
 limiting functions $E^{-1.5}$ and  $E^{-2}$. Labels denote the Fourier
 frequency with the $f<3$ denoting: 1, 0.6, 0.3 and 0.03 Hz. The frequencies
 in all panels are the same. 
 Panels {\it (b)\/} and {\it (c)\/} show the model of PF99 (i.e.\
 no radial dependence of flares' parameters, with spectral
 evolution during flares), for the PF99 and double exponential heating
 functions, respectively (see eq.~\ref{equ:pfheat} and \ref{equ:dbleexp}).
 Spectra were rescaled for clarity. Note opposite behaviour of the spectra 
 for the two $\lh(t)$, and no dependence on $f$ in the range $f>3$ Hz, where
 the PSD is formed by superposition of individual flares.
\label{fig:testffres}}
\end{figure*}

The spectrum of the reprocessed component is constructed according
to the simple formula of Lightman \& White (1988),
\begin{equation}
 \label{equ:lwrefl}
  \Sref(E) = {1-\epsilon \over 1+\epsilon} \Sprim(E), \quad\quad
      \epsilon = \sqrt{ {\kapabs \over \kapabs + \kapes}},
\end{equation}
where $\kapabs(E)$ and $\kapes(E)$ are the photo-absorption and electron
scattering opacities, respectively. ``Cold'' matter absorption opacities
and elements abundances from Morrison \& McCammon (1983) are used. 
Eq.~\ref{equ:lwrefl} gives the
angle averaged reflected spectrum and it is valid up to $\sim 15$ keV,
i.e.\ as long as the electron scattering can be treated as elastic.
Formula (\ref{equ:lwrefl}) is multiplied by a  simple exponential cutoff 
to mimic the proper shape of the reflected continuum at higher energies. 
This rough approximation to the reflected spectrum is sufficient for our
purposes, since the spectra only up to $\sim 30$ keV will be presented,
and it is not feasible to use e.g.\ the proper
angle-dependent functions of Magdziarz \& Zdziarski (1995).
The Fe \Ka line is added to the reprocessed continuum. Its equivalent
width (EW) is computed as a function of the spectral slope, $\Gamma$,
based on computations by \.{Z}ycki \& Czerny (1994). We note a number of
simplifications involved in the above treatment of the reprocessing:
most importantly, constant $\Gamma$ was used in computing $\tauhot$, 
while the continuum slope evolves during a flare. Furthermore, 
in our basic model {\em no\/} time delay 
is assumed in computing the reprocessed component, either due to
finite light travel time or adjustment of hydrostatic equilibrium
(Nayakshin \& Kazanas 2001). However, a possible influence of the
latter effect on the results will be presented in Sec.~\ref{sec:delay}
in a very simple approximation.

Finally, the $f$-resolved spectra are computed according to the
prescription given by Revnivtsev et al.\ (1999). The normalized power
spectral density (PSD) is computed as,
\begin{equation}
 P_j = 2 {T \over \bar{C}^2} |C_j|^2, \quad\quad j=0,\dots,N-1
\end{equation}
\begin{equation}
 C_j = \sum_{k=0}^{N-1}\,c_k\,e^{2\pi i f_j t_k},
\end{equation}
where the discrete frequencies $f_j\equiv j/T$ for $j=-N/2,\dots, N/2$;
$T$ is the total light curve time, $\bar{C}$ is the mean count rate,
while $c_k$ are the number of counts in $k$-th time bin. The $f$-resolved
spectrum at energy $E_i$ and Fourier frequency $f_j$ is then defined as
\begin{equation}
S(E_i, f_j) = \bar{C}_i\sqrt{P_i(f_j)\,\Delta f_j} = 
 \sqrt{ {2 |C_{i,j}|^2 \over T} \Delta f_j}.
\end{equation}

All computations are done for the central black hole mass $M=10\,\MSun$,
accretion rate $\Mdot = 0.01\MEdd$, where $\MEdd = \LEdd/(\varepsilon c^2)$
and $\varepsilon=0.057$ is the accretion efficiency. A fraction $\xi=0.2$
of total gravitational energy is assumed to be dissipated in the disc,
with the rest converted to hard X--rays (e.g.\ Di Salvo et al.\ 2001).

\section{Test runs}
\label{sec:test}

Before presenting the results of the full model computations let us
consider simpler test runs, which turn out to be helpful in understanding
the interpretation of the Fourier-$f$ resolved spectra.

The first, trivial test run is to assume that (1) there is no spectral 
evolution during a flare, i.e.\ the feedback between the soft and hard 
X--rays is switched off, and (2) all flares have the same spectra (no
radial dependence of any of the parameters). 
The reprocessed component is not computed.
All frequency-resolved spectra are then expected to be the same
and equal to the assumed spectrum,
as is indeed the case in our computations.

Second, we set up a run with an artificial dependence of spectrum on
the flare time scale, $\tau$, but still no spectral evolution during  flares.
Specifically, we assume  a power law $\Gamma(\tau)$ function such,
that the shortest flares (i.e.\ located at $r=\rmin$) have hard spectra,
$\Gamma = 1.5$, while the longest ones (at $r=\rmax$) have $\Gamma=2$.
The reprocessed component is not computed.
Result plotted in Figure~\ref{fig:testffres}a show, as expected, the high 
frequency spectra are hard and the spectra soften with decreasing $f$. 
Below $f\approx 3$ Hz, where PSD is formed by flares' avalanches, 
the slope of the energy spectra  is no longer dependent on $f$. 

Third, we introduce spectral evolution during flares, but neither
the radial dependences nor reprocessing.
The model is thus equivalent to that of PF99, with the same values of
all parameters, $\lambda=45$; $\mu=0.8$; $\alpha=7$; $p=1$; $\tmin=1$ ms;
$\tmax=0.15$ s, $H_0=0.5$, $D_0=0.5$. The frequency-resolved
spectra now too show some Fourier frequency dependence, even though all
flares have the same time evolution (in time units of $\tau$).  
In Fig.~\ref{fig:testffres}b,c the spectra are plotted for the two flare
profiles  (eq.~\ref{equ:pfheat} and eq.~\ref{equ:dbleexp}). The dependencies
on $f$ are opposite for the two profiles: for the PF99 profile
the spectra are harder at {\it lower\/} frequencies, while the opposite
is true for the double exponential profile. All $f$-resolved spectra are 
identical in the frequency range where individual flares form the PSD, 
i.e.\ $f>3$ Hz (see fig.~2b in PF99), the latter value being related to 
$\tmax$ and chosen to match the break in observed  PSD.

The opposite behaviour of energy spectra for $f<3$ Hz is related to 
 differences in spectral evolution during a flare for the two profiles.
For the PF99 profile the rise time at soft X-rays is shorter than
at hard X-rays (see fig.~1b in PF99), therefore the soft X-ray PSD breaks
at higher frequency than the hard X-ray PSD. The high-$f$ energy spectra are 
thus softer. The opposite is true for the double exponential profile:
the soft X-ray rise time is longer here, the soft X-ray PSD breaks at lower
$f$, and the higher-$f$ energy spectra are harder.

\begin{figure*}
 \parbox{\textwidth}{
 \hfil 
 \parbox{0.41\textwidth}{
 \epsfxsize 0.4\textwidth
 \epsfbox[18 200 580 710]{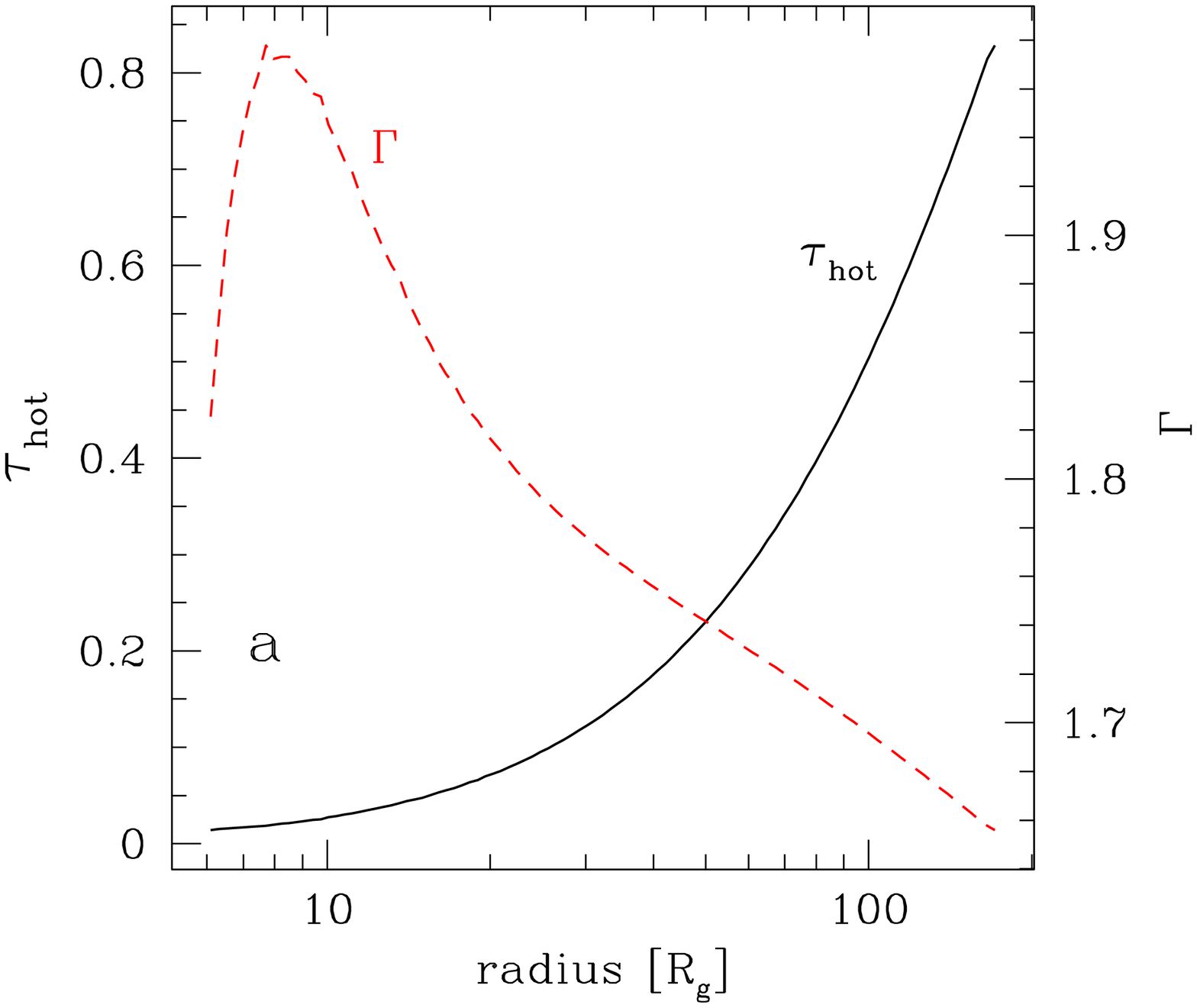}
 }  \hfil {
 \parbox{0.41\textwidth}{
 \epsfxsize 0.4\textwidth
 \epsfbox[18 200 580 710]{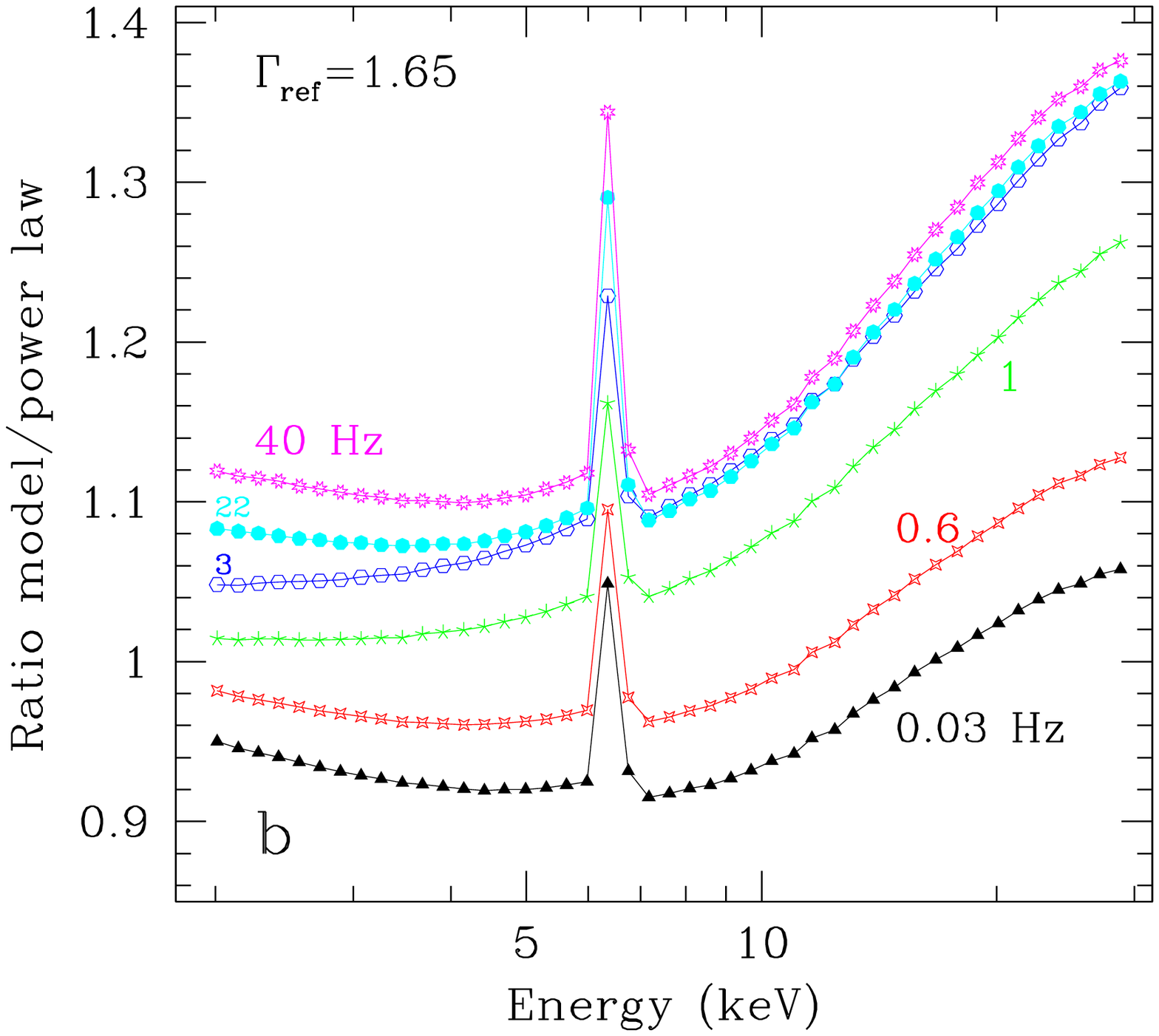}
}}\hfil}
 \parbox{\textwidth}{
 \hfil 
 \parbox{0.41\textwidth}{
 \epsfxsize 0.4\textwidth
 \epsfbox[18 190 580 690]{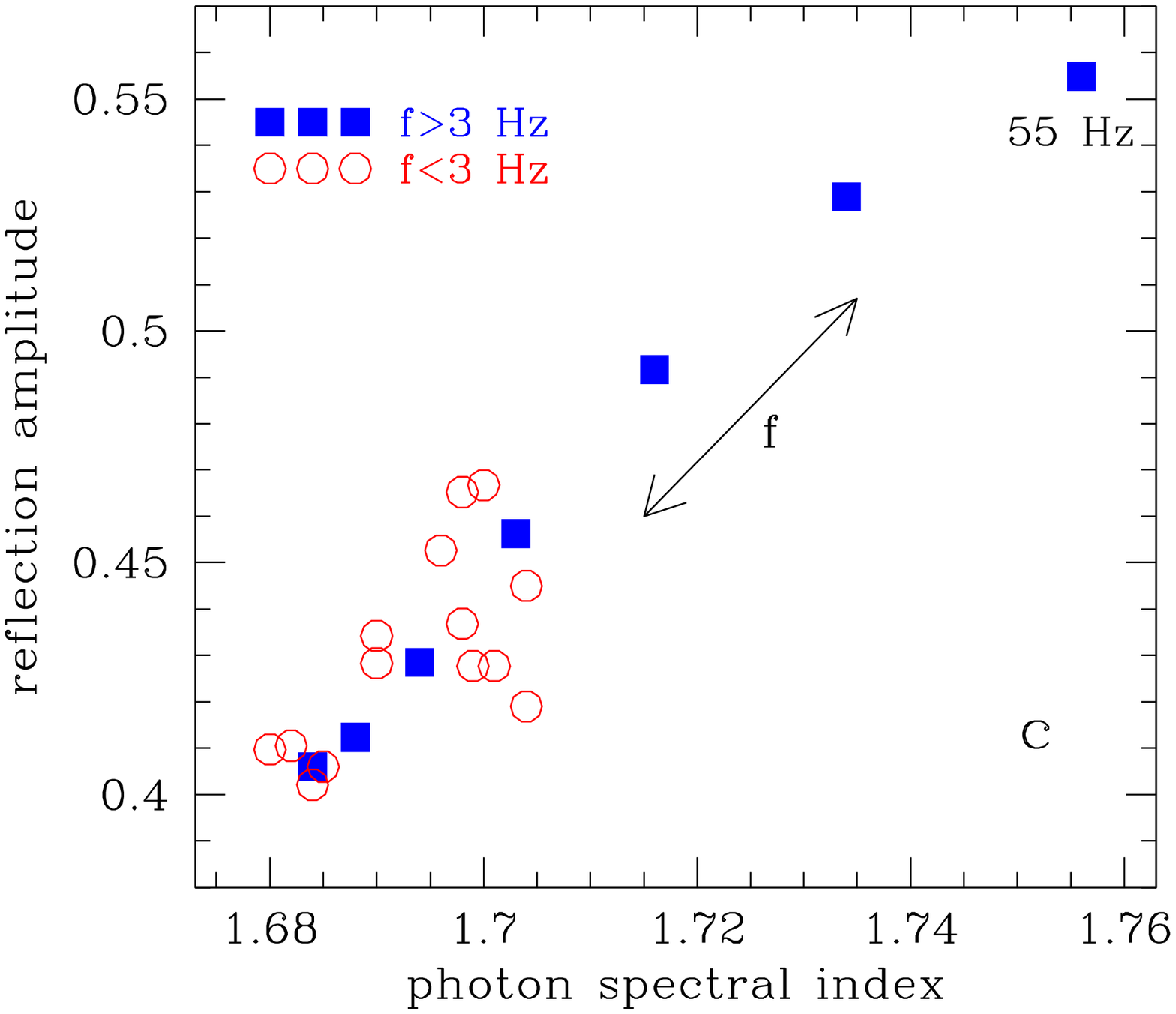}
 }  \hfil {
 \parbox{0.41\textwidth}{
 \epsfxsize 0.4\textwidth
 \epsfbox[18 190 580 690]{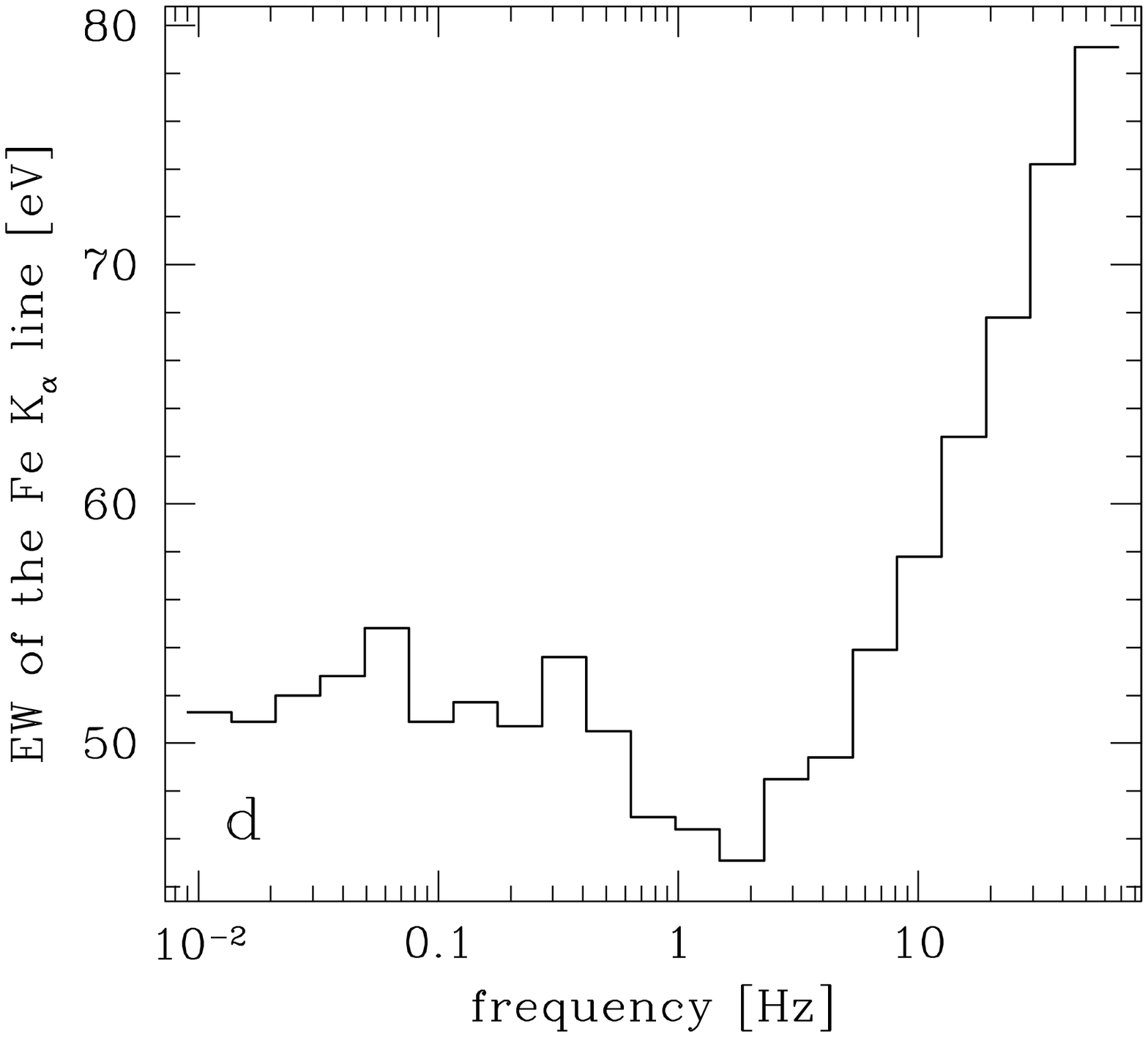}
}}\hfil}
 \caption{Results of computations for the basic model with
the double exponential $\lh(t)$ (eq.~\ref{equ:dbleexp}). Panel (a)
shows the flux-average $\tauhot$ and $\Gamma$ as functions of radius.
The hot skin is thickest at the {\it maximum\/} radius and, as a consequence,
the spectrum is hardest there. This means that the high-$f$ spectra, 
originating at small $r$, are soft, reversing the trend of hardening with $f$,
as shown in panel (b) -- compare with Fig.~\ref{fig:testffres}c.
The spectra plotted in panel (b) are for $f=0.03$, 0.6, 1, 3, 22 and 40 Hz,
ordered by increasing normalization at 2 keV.
Panel (c) plots the $R$--$\Gamma$ correlation with the Fourier frequency
 as the control parameter. The correlation holds only for high frequency, 
$f\ge 3$ Hz (solid squares), i.e.\ when the PSD is formed by individual flares, 
 but it breaks down for avalanches of flares ($f<3$ Hz, empty circles). 
The EW of the Fe \Ka line (panel d) shows a complicated
behaviour: it follows the $R$--$\Gamma$ correlation for $f\ge 3$ Hz; below
this frequency it is roughly constant, with a weak minimum $f\approx 1$ Hz.
\label{fig:full1}}
\end{figure*}

\section{Full model runs}
\label{sec:runs}

The main qualitative results of the computations with the hot skin 
can actually be easily
predicted. Assuming $\Rflare(r) = {\rm const}$ and $\lhmax(r)={\rm const}$
we obtain that the maximum illuminating flux is independent of radius.
Since the gravity decreases with radius, the thickness of the hot skin
will {\em increase\/} with radius. This is opposite to what is observed
in usual computations (Nayakshin 2000; \.{Z}ycki \& R\'{o}\.{z}a\'{n}ska
2001), where the illuminating flux is proportional to the
locally dissipated gravitational energy.
Moreover, the total luminosity of a flare is proportional to its
time scale, $\Lflare = \int_0^T \Lflare(t)\,dt \propto \tau$ for both
flare profiles. Thus the emitted radiation is dominated by
emission from outer radii (recall that by assumption $r \propto \tau^{2/3}$)
of the emitting region, i.e.\ where the hot skin is thickest.

\subsection{Basic model}

In our basic model we assume the double exponential profile of the flare 
heating function, eq.~\ref{equ:dbleexp}, with the following parameters:
$\lambda = 60$, $\mu=0.7$, $\alpha=5$, $p=1$, $\tmin=1$ ms, $\tmax=0.15$ s.
The value of $H_0$ ($\propto$vertical velocity of the flare) is such that
the height of the flare above the disc is equal to its radius at the
peak of energy dissipation, while $D_0=0.7$, so that the average
spectrum has slope $\Gamma\approx 1.7$.

The full numerical simulations confirm the above qualitative estimates. 
The flux-weighted average thickness of the hot skin is plotted in 
Figure~\ref{fig:full1}a, together with the average slope of the
Comptonized continuum. The thickness reaches maximum at the maximum
radius (recall that the existence of the maximum radius comes from
the assumption of a unique relation between the flare time scale
and radius). The Comptonized continuum originating at
larger distances is then indeed harder. Spectral index, $\Gamma$,
is smaller than in the test run (for the same values of other parameters) 
because of
the reduction of the soft reprocessed flux due to the hot skin.

The $f$-resolved spectra are presented in Figure~\ref{fig:full1}b.
The high frequency, $f\ge \fbreak$  ($\fbreak\approx 2$ Hz), 
spectra are now split due to the radial
dependence of $\tauhot$ (compare with Fig.~\ref{fig:testffres}c). 
Above $\fbreak$, the higher the frequency the softer (intrinsically)
the spectra and the larger the amplitude of the reprocessed component
(so the hardening above $\approx 10$ keV is more significant).
The $R$--$\Gamma$ correlation holds for these high Fourier frequencies,
as shown in Figure~\ref{fig:full1}c. 
The values of $R$ and $\Gamma$ plotted there at each $f$  were obtained
by constructing an e.g.\ ``$\Gamma$-spectrum'', 
$S(\Gamma, t) = \lh(t)\times G[\Gamma-\Gamma_0(t)]$ 
at each point in time, where $G(x)$ is a (narrow) gaussian function and
$\Gamma_0$ is the current value of the spectral index. The $\Gamma$-spectrum
thus describes the current distribution of the spectral index, analogously
to the energy spectrum describing the distribution of energy. Obviously,
since $\Gamma$ has a well defined value at each time step, its distribution
is strictly speaking a Dirac $\delta$-function. Thus obtained
sequence of spectra can be analyzed analogously to the usual energy
spectra. In particular $f$-resolved $\Gamma$-spectra were found, and the
mean values of $\Gamma$ and $R$ from these latter spectra are plotted
in Fig.~\ref{fig:full1}c. An alternative procedure of determining 
$\Gamma$ and $R$ for each $f$ would be to fit a reflection model to a 
spectrum for a particular Fourier component. The result would however be 
dependent on the reflection model 
used, hence the above, model independent procedure was adopted.
Variation of $\Gamma$ below $f\approx 3$ Hz is due to spectral evolution 
during a flare (Fig.~\ref{fig:testffres}) rather than the presence of the 
hot skin and, as a consequence, the $R$--$\Gamma$ correlation does not
hold for those lower frequencies. This can also be seen from 
Fig.~\ref{fig:full1}d, where the EW of the Fe \Ka line is plotted
as a function of $f$. There is a clear monotonic dependence EW$(f)$
for $f\ge \fbreak$, but not so below this frequency. 

\begin{figure}
 \epsfxsize 0.45\textwidth
 \epsfbox[18 250 580 700]{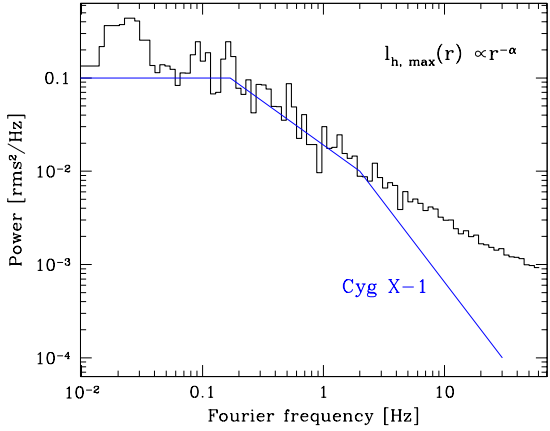}
 \caption{PSD from the 3 keV lightcurve in the model with radial dependence of
 $\lhmax$: $\lhmax(r) \propto r^{-2.5}$. The diameter
 of the flare is assumed constant.  The radial dependence of the peak flare 
 luminosity influences the distribution of number
 of events over luminosity and, consequently, affects the PSD. 
 In this example the PSD has no high-$f$ break, which is commonly
 observed in black hole binary systems at $f\sim 1$ Hz.
\label{fig:lhmax}}
\end{figure}

\subsection{Comparison with observations}

The trends obtained in the presence of the hot skin are clearly opposite to 
what was found from observations
of black hole binaries. For both Cyg X--1 and GX~339-4 the $f$-resolved
spectra are harder when $f$ increases (Revnivtsev et al.\ 1999, 2001). This
holds best for the frequency range $\approx 0.5$--$12$ Hz,  where both 
single flares and flares' avalanches form the PSD. The $\Gamma(f)$ dependence
seems to flatten both at low, $f<0.5$ Hz, and high, $f>10$ Hz, frequencies
(Revnivtsev et al.\ 1999). Similar conclusion,
that the shots contributing to low-$f$ part of PSD have softer spectra
than average, was reached by  Negoro, Kitamoto \& Mineshige (2001), using
the technique of superposed shots profile.
The EW of the Fe \Ka
line decreases monotonically with $f$ for $f>0.5$ Hz, being $\sim$constant
below that frequency (Revnivtsev et al.\ 1999, 2001). Again this is
incompatible with the model, where a {\it rise\/} of EW with $f$, for
$f>1$ Hz is observed. At low frequency, $f<1$ Hz, the model EW$(f)$ is 
$\sim$constant at $\approx 40$ eV, i.e.\ the reflection is least efficient.
The observed value is 
$\approx 100$ eV, compatible with amplitude of reflection $R\approx 1$
(Revnivtsev et al.\ 1999).

\subsection{Variations on the basic model}
 \label{sec:varia}

\subsubsection{Radial dependence of the flare peak luminosity}
 \label{sec:radiallheat}

From the discussion above it is clear that the opposite behaviour of
Comptonized spectra with Fourier frequency to that observed is partly 
due to the peak flares luminosity being constant with radius. Additionaly, 
the increase of luminosity with radius is obviously in 
disagreement with the radial dependence of the gravitational energy release, 
$dL/dr(r) \propto r r^{-3}$. It might therefore seem appropriate to modify the
$\lhmax(r)$ dependence assuming it should decrease with radius, so that
the longer flares have lower peak luminosities.
However, any such modification will affect the PSD. This is because the PSD
is related to the distribution of the number of events (individual flares and 
avalanches) over their (total) luminosity. 
The $f^{-1}$ behaviour is a direct result of superposing many small events,
fewer bigger events, still fewer even bigger events, etc.\ (e.g.\ Bak 1997).
The break at $f\approx 2$ Hz in PSD in the 
original model of PF99 seems to be related to a break in the distribution
$dN/dL$ in this model which occurs at 
$L = \tmax \lhmax \times 2\Rflare \me c^2/\sigmaT$ (i.e.\ total luminosity of
longest flares).
Decreasing $\lhmax$ at the maximum radius (i.e.\ longest flares)
decreases the range of $L$ spanned by individual flares. The resulting
distribution  has no clear break and, as a result, the PSD has no break
in the whole range of frequencies of interest (Fig.~\ref{fig:lhmax}).
This is obviously incompatible with the observed PSD of X--ray lightcurves,
which universally show breaks at a few Hz. The conclusion about lack
of break in PSD seems to be independent of the value of the exponent
in the $P(\tau)$ probability (see Sec.~\ref{sec:tvaria}), provided that
$\lhmax(r)$  is adjusted to give an assumed $dL/dr(r)$ 
(e.g.\ $\propto r^{-2}$)  dependence.
Since the total emission from each flare is
$\propto \tau\lhmax$,  the radial distribution of emission is
$dL/dr \propto dN/d\tau \times \tau\lhmax \propto 
P(\tau) \tau \lhmax$. For the assumed $P(\tau) \propto \tau^{-1}$
and $\tau(r)\propto r^{3/2}$ (Eq.~\ref{equ:rvstau}),
the distribution becomes $dL/dr \propto r^{1/2} \lhmax(r)$.
The PSD plotted in Fig.~\ref{fig:lhmax} was computed assuming 
$\lhmax(r) \propto r^{-2.5}$, which gives $dL/dr(r) \propto r^{-2}$.

The effect of the hot skin in this model is insignificant. 
This is because, adjusting the normalization of $\lhmax(r)$  so that the
mean luminosity-weighted $\lh\approx 10$, the simple time average of $\lh$ is
much smaller than before. This requires larger sizes of flares, $\Rflare$, 
in order for the total luminosity to remain constant. 
Hence, the maximum illuminating flux, which scales as 
$F \propto L_{\rm flare}/h^2 
\propto 1/\Rflare^2$ (assuming the height of maximum dissipation
$h\approx \Rflare$), is smaller than in the model with $\lhmax(r)={\rm const}$.
As a consequence, no correlations is obtained between $R$ and $\Gamma$.
The $f$-resolved spectra in this model show slight hardening with $f$, due to 
flare spectral evolution (see Fig.~\ref{fig:testffres}c), but
the reprocessed component is constant with $f$.

\begin{figure}
 \epsfysize 7cm
 \epsfbox[18 220 600 700]{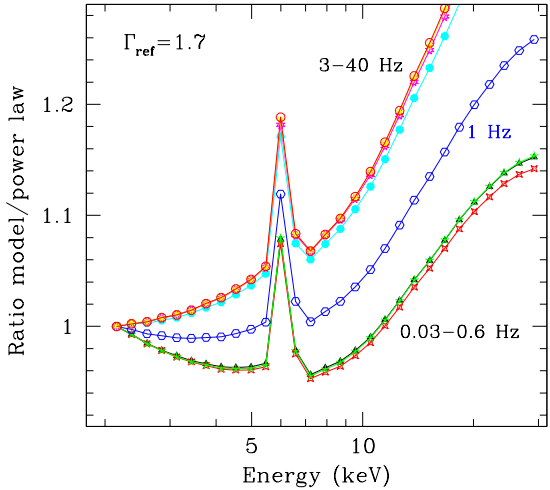}
 \caption{Fourier $f$-resolved spectra in the modified model, assuming
 that the location of {\em stimulated\/} flares is the same as their 
 stimulating  flare  (Sec.~\ref{sec:taurad}), 
 i.e.\ whole avalanche developes in the  same place. The high-$f$ ($f>3$ Hz)
 spectra are now the same, since the correspondence of time scale and
 physical conditions is lost. 
 \label{fig:ffresvar}}
\end{figure}

\subsubsection{Modified radial distribution of the stimulated flares}
  \label{sec:taurad}

The Fourier frequency dependence of the spectra becomes less significant,
when the assumption is made that the spatial positions of all stimulated flares
coincide with that of their stimulating flare (Sec~\ref{sec:radial};
the distribution of time scales $P(\tau)\propto \tau^{-1}$ is retained
for all flares). 
This is to be expected, since the correspondence of variability time scale
(i.e.\ $f$) and physical conditions (i.e.\ $r$) is  no longer unique.
In particular, the high-$f$ spectra are almost identical in this case,
while the lower-$f$ ones are somewhat softer compared to results of the
basic model (Fig.~\ref{fig:ffresvar}). The EW of the \Ka line
shows no frequency dependence being approximately constant at $\approx 60$ eV.

\subsubsection{Delayed response of the reflecting medium}
 \label{sec:delay}

The properties of the reprocessed component may not be always given by 
the instanteneous illuminating flux, as the accretion disc needs to re-adjust
to the condition of hydrostatic equilibrium with changed X--ray radiation
pressure and the Compton temperature (Nayakshin \& Kazanas 2001).  
Parameters of the hot skin are thus
not given by the instantaneous flux, if the flux fluctuates rapidly.
The thickness of the hot skin re-adjusts on the dynamical time scale
$t_{\rm response} \approx \tdyn = \OmegaK^{-1}$, while its temperature on 
the thermal time scale, which is usually much shorter (Nayakshin \& Kazanas 
2001). This effect may influence our results, since the response time
is approximately equal the flare time scale, $\tau \approx \tdyn
\approx t_{\rm response}$. Since proper computations
of time-dependent response of an accretion disc to changing illumination
are not possible at present, we  model here the response in a rather 
approximate way, motivated by the results of Nayakshin \& Kazanas (2001).
We simply assume that for each flare $\tauhot(t)$ is determined by
illumination flux at an eariler time $t_1 = t - t_{\rm response} = 
t - \tdyn$, i.e.\ the complex response of reflection is approximated as 
a simple time shift. In particular, at the beginning of a flare, for
$t<\tdyn$,  there is no hot skin and the reprocessed component is produced
with amplitude 1.

The radial dependence of $\tauhot$ in this model is qualitatively the
same as in the basic model, i.e.\ $\tauhot$ increases with $r$.
The values of $\tauhot(r)$ are now smaller, though, since by assumption
$\tauhot$ is fixed at zero for a certain time $\approx \tau$ during each flare.
The radial dependence of $\Gamma$ is somewhat weaker but the overall
trends are virtually the same as in our basic model (Fig.~\ref{fig:full1}b). 
With increasing $f$
the spectra become harder first, then, above $f\approx 3$ Hz, they soften.
The EW of the \Ka line has the same $f$-dependence except for a slight
overall increase by 15--20 eV.  There results are not unexpected since 
the delay is relatively short compared to the duration of a flare. With 
increasing delay, still the same qualitative results are obtained. 
The influence of
the hot skin becomes less significant but the radial trends are the same, if
the delay time scale is related to the flare time scale. In the limit of a very
long delay the presence of the hot skin becomes unimportant, and the model
reduces to the original model of PF99.

\section{Discussion}
\label{sec:discuss}

A number of geometrical/physical models of accretion onto compact sources 
can be constructed, if only a part of the available information on X-ray 
emission is used. Supplementing the usually employed energy spectral
analysis with time variability and, in particular, correlated 
spectral--timing behaviour, one can attempt to distinguish between
the models. In this paper predictions for Fourier-resolved spectroscopy
are presented for the model of magnetic flares avalanches (PF99) above
an accretion disc with hot ionized skin.

The basic model formulated by PF99 is quite successful in explaining
both energy spectra and a number of timing characteristics: PSD,
hard X--ray time lags and the coherence function. With the suitably
chosen flare profile (e.g.\ a double exponential with $t_{\rm rise}\gg
t_{\rm decay}$), the auto- and cross-correlation function can also
be reproduced (Maccarone et al.\ 2000). The dependence of energy
spectra on Fourier frequency, resulting from the spectral evolution during
flares alone, is in agreement with observations for e.g.\ the double 
exponential flare profile, but not for the profile assumed in PF99
(Sec~\ref{sec:test}).

Our extension of the PF99 model is to assume that there exists at least
partial correspondence between location of a flare and its time scale.
The simplest assumption about such correspondence
is that it follows from projecting the flares time scales onto 
the (inverse of) Keplerian  frequency. 
The shortest flares are then located near the inner radius of the disc, 
the longer a flare the farther away it is located. 
Now, in the basic model both the peak heating rate (compactness parameter)
and flare diameter are independent of radius, which means that the
peak flux of the disc illuminating X--rays is independent of radius, too. 
Since the vertical
component of gravity decreases with radius ($g_{\rm z} \propto (z/r) r^{-2}$),
the thickness of the hot ionized skin {\em increases\/} with radius.
The fast flares, located where the hot skin is thinest, have then softer
spectra than the slower flares located where the thick hot skin
decreases the effectiveness of thermalization of X--rays and,
consequently, the flux of soft photons for Comptonization.
As a result, the energy spectra corresponding to high Fourier
frequencies are {\em softer\/} than the lower $f$ spectra, for the
frequency where the properties of individual flares determine
the PSD, $f>\fbreak$, where $\fbreak=1$--3 Hz in black hole binaries. 
This is opposite to the observed trends,
which show steady hardening of the spectra with increasing Fourier
frequency, at least up to $\approx 12$ Hz (Revnivtsev et al.\ 1999).

As a consequence, the properties of the reprocessed component in the 
presented model, 
in particular the Fe \Ka line, do not agree with the observed trends,
either. At high $f>\fbreak$ the  reflection amplitude and  
EW of the \Ka line increase sharply with $f$, as a consequence of 
the inwards--decreasing thickness of the hot skin.
At low $f<1$ Hz the model $R$ and EW of the line are roughly constant,
however $R$ is significantly less than 1 and, consequently, EW is smaller
than the corresponding value of $\sim 100$ eV. This is contrary to the
observed $R\approx 1$ and EW$\approx 100$ eV at these low frequencies.

These results hold true even when some of the model assumptions are
modified (Sec.~\ref{sec:varia}). None of the modifications made:
the different radial distribution of the stimulated flares, 
radial dependence of the
peak flare luminosity or the delayed response of reflection, changed
the trends  of spectral parameters with Fourier frequency.
It is worth emphasizing that the perhaps most obvious of the above 
modifications -- the radial distribution of flares peak luminosity such that 
the total flares luminosity decreases with radius -- not only does not help 
to reproduce the observed EW$(f)$ and $\Gamma(f)$, 
but it also changes the PSD so that it no longer agrees with observations.
It remains to be seen whether a more complex formulation of the model is 
possible, which would simultaneously predict the correct PSD and
thickness of the hot skin decreasing with radius.

The same technique of analysis of $f$-resolved spectra can be applied to 
other models mentioned in 
Sec.~\ref{sec:intro}. In the dynamic corona model with bulk outflow
of the plasma, the plasma velocity, $\beta$, is the control parameter of the
$R$--$\Gamma$ correlation (Beloborodov 1999a,b). 
The observed $f$-dependence of the energy
spectra would require  higher $\beta$ closer to the center. 
This is indeed expected, if the X-ray luminosity scales with local
gravitational energy dissipation. However, if the X--ray emission is determined
from the properties of the flares, assuming their radial distribution as
adopted in this paper (Sec.~\ref{sec:radial}, Eq.~\ref{equ:rvstau}), 
then stronger X--ray emission is expected from larger distances, exactly
as in the computations of the hot skin model.
It is then likely that $\beta$ will increase with radius, analogously
to the result that the thickness of the hot skin increases with radius
in our model. The result for the $f$-resolved spectra will then be the same
as for the model with hot skin.
In the alternative geometry (the optically thick
flow distrupted at certain radius and replaced with some form of hot optically
thin flow), the model of B\"{o}ttcher \& Liang (1999) can be used to compute
the $f$-resolved spectra. Alternatively, the idea of Kotov, Churazov \& 
Gilfanov (2001; see also Nowak et al.\ 1999), 
that the X--ray emitting perturbations are propagating
inwards, if developed further into a more physical model, could be used
to predict the $f$-resolved spectra.

\section{Conclusions}

\begin{itemize}
 \item Fourier-resolved X--ray spectroscopy is a  powerful tool for
  discriminating various models of accretion.
 \item The model of (avalanches of) magnetic flares, radially distributed 
  above an accretion disc with hot  
  ionized skin,  predicts dependencies of various spectral parameters
  ($\Gamma$, $R$, EW of the Fe \Ka line) on Fourier frequency
  opposite to those observed.
\end{itemize}

\section*{Acknowledgments} 
 
I acknowledge informative discussions with Bo\.{z}ena Czerny, Juri Poutanen 
and Lev Titarchuk on various aspects of time variability, and a fruitful
collaboration with Agata R\'{o}\.{z}a\'{n}ska on illuminated accretion discs.
The referee, Juri Poutanen, made a number of comments which improved
the presentation of the results.
This work  was partly supported by grant no.\  2P03D01718 
of the Polish State Committee for Scientific Research (KBN).

{}



\begin{thebibliography}{}

 \bibitem[]{}
   Bak P., 1997, How Nature works, Oxford University Press, Oxford.
 \bibitem[]{}
   Beloborodov A. M., 1999a, ApJ, 510, L123
 \bibitem[]{}
   Beloborodov A. M., 1999b,  in Poutanen J., Svensson R. eds, 
     ASP Conf. Ser. 161, High Energy Processes in Accreting Black Holes, 
     295,  (astro-ph/9901108)
 \bibitem[]{}
   B\"{o}ttcher M., Liang E. P., 1999, ApJ, 511, L37
 \bibitem[]{}
  Collin-Souffrin S., Czerny B., Dumont A.-M., \.{Z}ycki P. T., 1996, A\&A, 
                                  314, 393
 \bibitem[]{}
   Coppi P. S., 1999, in Poutanen J., Svensson R., eds, ASP Conf Ser. Vol. 161.
   Astron. Soc. Pac., San Francisco, p.~375 (astro-ph/9903158)
 \bibitem{}
   Di Salvo T., Done C., \.{Z}ycki P. T., Burderi L., Robba N. R., 2001, ApJ,
    547, 1024
 \bibitem[]{}
   Esin A. A., McClintock J. E., Narayan R., 1997, ApJ, 489, 865
 \bibitem[]{}
   Field G. B., 1965, ApJ, 142, 531
\bibitem []{}
  Gilfanov M., Churazov E.,  Revnivtsev M., 1999, A\&A, 352, 182
\bibitem []{}
  Gilfanov M., Churazov E.,  Revnivtsev M., 2000a, in Proc.\ of the 5th 
   Sino-German Workshop on Astrophysics, SGSC Conference Ser., Vol. 1,
  China Science \& Technology Press, Beijing, p.~114 (astro-ph/0002415)
\bibitem []{}
  Gilfanov M., Churazov E.,  Revnivtsev M., 2000b, MNRAS, 316, 923
 \bibitem[]{}
   Janiuk A., Czerny B., \.{Z}ycki P., 2000, MNRAS, 318, 180
 \bibitem[]{}
   Kazanas D., Hua X. M., Titarchuk L., 1997, ApJ, 480, 735
 \bibitem[]{}
   Kotov O., Churazov E., Gilfanov M., 2001, MNRAS, 327, 799
\bibitem[]{}
   Krolik J. H., McKee C. F., Tarter C. B., 1981, ApJ, 249, 422
 \bibitem[]{}
   Kuncic Z., Celotti A., Rees M. J., 1997, MNRAS, 284, 717
 \bibitem[]{}
   Lehto H. J., 1989, in The 23rd ESLAB Symposium on Two Topics in X--Ray 
   Astronomy. Volume 1: X Ray Binaries, ESA, p.~499
 \bibitem[]{}
   Lightman A. P., White T. R., 1988, ApJ, 335, L57
 \bibitem[]{}
   Lochner J. C., Swank J. H., Szymkowiak A. E., 1991 ApJ, 376, 295
 \bibitem[]{}
   Lubi\'{n}ski P., Zdziarski A. A., 2001, MNRAS, 323, L37
 \bibitem[]{}
   Maccarone T. J., Coppi P. S., Poutanen J., 2000, ApJ, 537, L107
 \bibitem[]{}
    Magdziarz P., Zdziarski A. A., 1995, MNRAS, 273, 837
 \bibitem[]{}
   Malzac J., 2001, MNRAS, 325, 1625
 \bibitem[]{}
   Malzac J., Beloborodov A. M., Poutanen J., 2001, MNRAS, 326, 417
 \bibitem[]{}
   Merloni A., Fabian A. C., 2001, MNRAS, 328, 958
 \bibitem[]{}
   Miyamoto S., Kitamoto S., 1989, Nat, 342, 773
 \bibitem[]{}
   Morrison R.,  McCammon D., 1983, ApJ, 270, 119
 \bibitem[]{}
   Nayakshin S., 2000, ApJ, 534, 718
 \bibitem[]{}
   Nayakshin S., Kazanas D., 2001, ApJ, in press (astro-ph/0106450)
 \bibitem[]{}
   Nayakshin S., Kazanas D., Kallman T. R., 2000, ApJ, 537, 833
  \bibitem[]{}
    Negoro H., Kitamoto S,  Mineshige S., 2001, ApJ, 554, 528
 \bibitem[]{}
   Nowak M. A., Wilms J., Vaughan B. A., Dove J. B, Begelman M. C., 1999, 
             ApJ, 515, 726
 \bibitem[]{}
   Poutanen J., 1998, in Abramowicz M. A., Bj\"{o}rnsson G., Pringle J. E., 
   eds, Theory of Black Hole Accretion Discs. CUP, Cambridge, p.~100
   (astro-ph/9805025)
 \bibitem[]{}
   Poutanen J., 2001, AdSpR, 28, 267 (astro-ph/0102325)
 \bibitem[]{}
   Poutanen J., 2002, MNRAS, in press (astro-ph/0106189)
 \bibitem[]{}
   Poutanen J., Fabian A. C., 1999, MNRAS, 306, L31 (PF99)
\bibitem[]{}
   Poutanen J., Krolik J. H., Ryde F., 1997, MNRAS, 292, L21
\bibitem[]{}
   Rees M., 1987, MNRAS, 228, 47P
 \bibitem[]{}
   Revnivtsev M., Gilfanov M., Churazov E., 1999, A\&A, 347, L23
 \bibitem[]{}
   Revnivtsev M., Gilfanov M., Churazov E., 2001, A\&A, 380, 520
 \bibitem[]{}
   R\'{o}\.{z}a\'{n}ska A., 1999, MNRAS, 308, 751
 \bibitem[]{}
   R\'{o}\.{z}a\'{n}ska A., Czerny B., 1996, Acta Astron., 46, 233 
 \bibitem[]{}
   R\'{o}\.{z}a\'{n}ska A., Czerny B., 2000, A\&A, 360, 1170
 \bibitem[]{}
   Shakura N. I., Sunyaev R. A., 1973, A\&A, 24, 337
 \bibitem[]{}
   Stern B. E., Svensson R., 1996, ApJ, 469, L109
 \bibitem[]{}
   Zdziarski A. A., Johnson W. N., Magdziarz P., 1996 ApJ, 283, 193
 \bibitem[]{}
   Zdziarski A. A., Lubi\'{n}ski P., Smith D. A., 1999, MNRAS, 303, L11
 \bibitem[]{}
    \.{Z}ycki P. T., Czerny B., 1994, MNRAS, 266, 653
 \bibitem[]{}
    \.{Z}ycki P. T., R\'{o}\.{z}a\'{n}ska A., 2001, MNRAS, 325, 197
 \bibitem[]{}
    \.{Z}ycki P. T., Done C.,  Smith D. A., 1998, ApJ, 496, L25 

\label{lastpage}

\end{thebibliography}
\end{document}